\documentclass[aip,graphicx]{revtex4-1}
\usepackage[T2A]{fontenc}
\usepackage[english]{babel}
\usepackage{amssymb,latexsym,amsmath,amscd}
\usepackage{graphicx,color,framed}
\begin{document}
\selectlanguage{english}
\title{A new equation of state of a flexible-chain polyelectrolyte solution: Phase equilibria 
and osmotic pressure in the salt-free case}
\author{\firstname{Yu.~A.} \surname{Budkov}}
\email[]{urabudkov@rambler.ru}
\affiliation{G.A. Krestov Institute of Solution Chemistry of the Russian Academy of Sciences, Ivanovo, Russia}
\affiliation{National Research University Higher School of Economics, Department of Applied Mathematics, Moscow, Russia}

\author{\firstname{ A.~L.} \surname{Kolesnikov}}
\affiliation{Ivanovo State University, Ivanovo, Russia}
\affiliation{Institut f\"{u}r Nichtklassische Chemie e.V., Universitat Leipzig, Leipzig, Germany}

\author{\firstname{ N.} \surname{Georgi}}
\affiliation{Max Planck Institute for Mathematics in the Sciences, Leipzig, Germany}
\author{\firstname{E.~A.} \surname{Nogovitsyn}}
\affiliation{Ivanovo State University, Ivanovo, Russia}

\author{\firstname{M.~G.} \surname{Kiselev}}
\affiliation{G.A. Krestov Institute of Solution Chemistry of the Russian Academy of Sciences, Ivanovo, Russia}
\begin{abstract}
We develop a first-principle equation of state of salt-free polyelectrolyte solution in the limit of
infinitely long flexible polymer chains in the framework of a field-theoretical formalism 
beyond the linear Debye-Hueckel theory and predict a liquid-liquid phase separation induced 
by a strong correlation attraction. As a reference system we choose a set of two subsystems -- 
charged macromolecules immersed in a structureless oppositely charged 
background created by counterions (polymer one component plasma) and counterions immersed in 
oppositely charged background created by polymer chains (hard-core one component plasma). 
We calculate the excess free energy of polymer one component plasma in the framework 
of Modified Random Phase Approximation, whereas a contribution of charge densities 
fluctuations of neutralizing backgrounds we evaluate at the level of Gaussian approximation. 
We show that our theory is in a very good agreement with the results of Monte-Carlo and MD simulations 
for critical parameters of liquid-liquid phase separation and osmotic pressure in 
a wide range of monomer concentration above the critical point, respectively.
\end{abstract}
\maketitle

\section{Introduction}
It is well known that thermodynamic and structural properties of polyelectrolyte in homogeneous solutions are mainly determined by the correlation attraction of like-charged particles which are related to long-range Coulomb interactions \cite{Frederickson,Holm_Review,Borue_1988,Muthu_1996,Levin,Winkler,Joanny_1990,Cherstvy,DeLaCruz_2013,Baeurle_2009}. Moreover the collective effects in semi-dilute regime of polyelectrolyte solutions resulting from strong fluctuations of monomer concentration, due to the chain structure, play a crucial role \cite{Khohlov,DeGennes}. The necessity to take into account the two effects simultaneously complicates the theoretical description of the thermodynamics of polyelectrolyte solution in a wide range of monomer concentration and temperature.

It is well known that in salt-free flexible chain polyelectrolyte solutions a liquid-liquid phase separation 
due to strong correlation attraction can take place. This effect has been described in several theoretical works 
\cite{Kramarenko_2002,Warren_1997,Mahdi_2000,Ermoshkin,Muthu_2002,Muthu_2009,JIANG_2001} and has also been shown 
in Monte-Carlo simulations \cite{Kumar_PRL}. However, to the best of our knowledge theoretical models correctly 
predicting critical parameters of such a purely electrostatics driven phase transition are still missing. 
The reason why theoretical predictions of the critical parameters significantly deviate from results of 
Monte-Carlo simulation are three-fold.

Firstly, in a series of studies the contribution of the correlation attraction to the total 
Helmholtz free energy has been omitted \cite{Warren_1997,Khohlov_1992,Gottschalk_1998} or described at the level of Debye-Hueckel approximation \cite{Muthu_2002,Muthu_2009,Kramarenko_2002,Mahdi_2000}, although the critical point of the 
electrostatic phase transition is located far from the range of applicability of a linear Debye-Hueckel 
theory \cite{Kumar_PRL}. In work \cite{JIANG_2001} the contribution of the correlation attraction was 
described at the level of mean-spherical approximation (MSA) that allowed to obtain a good agreement 
with simulation results for critical temperature but underestimated the critical monomer concentration.
Secondly, most theoretical approaches \cite{Kramarenko_2002,Warren_1997,Mahdi_2000,
Ermoshkin,Muthu_2002,Muthu_2009} ignore the effects of the monomer concentration 
fluctuations in the regime of semi-dilute polymer solution. However, these effects, as was already mentioned above, 
should be important for solutions of polyelectrolytes as well as for neutral polymers \cite{Borue_1988,Frederickson}.
In order to evaluate the contribution of the non-electrostatic interactions to the total free energy of the solution 
the  authors usually resort to some mean-field type approximations \cite{Kramarenko_2002,Warren_1997,
Mahdi_2000,Ermoshkin,Muthu_2002,Muthu_2009} where the fluctuation effects are fully ignored 
\cite{Khohlov,DeGennes,Frederickson}. Finally, all the above mentioned theoretical models are based on the implicit assumption that contributions of electrostatic and non-electrostatic interactions are independent. 
We would like to stress that a methodology based on a $"$construction$"$ of the total free energy using expressions obtained from different theoretical approaches cannot be strictly justified from first principles of statistical thermodynamics.

In reference \cite{Ermoshkin} addressed an interesting observation that one can obtain reasonable 
values of the critical parameters if one uses the modified random phase approximation (MRPA) for the
calculation of the electrostatic contribution into the free energy of the solution. MRPA was developed in works 
\cite{Brilliantov_OCP,Brilliantov_HCOCP} and in contrast to the standard random phase approximation (RPA) 
contains a concept of ultraviolet cut-off in the procedure of integration over vectors of reciprocal space. 
The parameter of ultraviolet cut-off determined from the condition of equality between number of degrees 
of freedom of the system and number of collective variables which contribute into the total free energy
enumerated by elements of reciprocal space. Therefore, the cut-off parameter within MRPA is related 
to the number density of particles. It is surprising that MRPA allows one to obtain very accurate 
interpolation formulas for excess free energy of one component plasma (OCP) 
and hard-core one component plasma (HCOCP) in a wide range of temperatures and number densities -- 
from regime of linear Debye-Hueckel theory to the strong coupling regime \cite{Brilliantov_OCP,Brilliantov_HCOCP}.

In work \cite{Edwards_Muthu} in the framework of a field-theoretical approach the interpolation formulas for excess free energy, correlation length, and effective Kuhn length of the segment of polymer chain for semi-dilute solution of neutral polymer chains were obtained. It should be emphasized, that the interpolation formulas obtained by authors contain all limiting laws predicted in earlier works within the scaling approach \cite{DeGennes}. 
Using the widely used Carnahan-Starling interpolation formula for the free energy of a hard sphere system and 
the interpolation formula for excess free energy of semi-dilute polymer solution mentioned above \cite{Edwards_Muthu} 
offer now new opportunities for theoretical description of more complicated polymer systems 
in the framework of thermodynamic perturbation theory (TPT).

The presented observation in reference \cite{Ermoshkin}  and results proposed in reference \cite{Edwards_Muthu} 
motivated our attempt to develop a statistical theory of salt-free polyelectrolyte solutions which 
would simultaneously take into account the effect of correlation attractions of charged monomers 
and counterions beyond the linear Debye-Hueckel theory as well as the collective effects related to the 
monomer concentration fluctuations in the semi-dilute regime.

We formulate the theoretical model based on a variant of TPT formalism, 
which has been developed in works \cite{Budkov_2013_1,Budkov_2014}. As the reference 
system we have chosen a set of two independent subsystems -- charged polymer chains immersed 
in a structureless oppositely charged background created by counterions 
(polymer one component plasma) and counterions (which we model as charged hard spheres) 
immersed in oppositely charged background created by polymer chains (hard-core one component plasma). 
We calculate the excess free energy of the polymer one component plasma in the framework of MRPA. 
We show that our theory is in a very good agreement with Monte-Carlo simulations results
reproducing critical parameters of the electrostatic phase transition and MD simulation results reproducing 
the osmotic pressure as a function of monomer concentration in a region that is above the critical point, respectively.

\section{Theory}
\subsection{Hamiltonian}
In this section we present the theoretical model based on the TPT formalism for the thermodynamics 
of a salt-free flexible chain polyelectrolyte solution. 
The system contains charged polymer chain and counterions, where we assume that the charged polymer chain is immersed in a good solvent modelled by a structureless dielectric medium with the dielectric permittivity $\varepsilon$. 

Moreover, we take into account an excluded volume of monomers and counterions modeling them as hard spheres of diameter $d$. 
For simplicity we also assume that each monomer has a fixed charge $-e$, whereas the counterions carry an opposite charge $+e$ ( monovalent counterions). Thus the free energy of the solution  within the canonical ensemble is
\begin{equation}
F=-k_{B}T\ln{Z},
\end{equation}
where the partition function $Z$ takes the form
\begin{equation}
\label{eq:Z}
Z=\frac{1}{N_{p}!N_{c}!}\int d\Gamma_{p}\int d\Gamma_{c} e^{-\beta H},
\end{equation}
with
\begin{equation}
\label{eq:H}
H=H^{(c)}_{e.v.}+H^{(p)}_{e.v.}+H^{(pc)}_{e.v.}+\frac{1}{2}\left(\hat{\rho}_{p}V_{c}\hat{\rho}_{p}\right)+\frac{1}{2}\left(\hat{\rho}_{c}V_{c}\hat{\rho}_{c}\right)+\left(\hat{\rho}_{p}V_{c}\hat{\rho}_{c}\right)\nonumber
\end{equation}
being the Hamiltonian of the system, where the following short-hand notations have been introduced
\begin{equation}
\label{eq:fVcg}
\left(f V_{c}g\right)=\int d\bold{x}\int d\bold{y}f(\bold{x})V_{c}(\bold{x}-\bold{y})g(\bold{y}),
\end{equation}
\begin{equation}
\label{eq:Vc}
V_{c}(\bold{x}-\bold{y})=\frac{1}{\varepsilon |\bold{x}-\bold{y}|}
\end{equation}
is the Coulomb potential,
\begin{equation}
\hat{\rho}_{p}(\bold{x})=-e\sum\limits_{i=1}^{N_{p}}\int\limits_{0}^{N}ds\delta (\bold{x}-\bold{R}_{i}(s))
\end{equation}
is the microscopic charge density of the monomers,
\begin{equation}
\hat{\rho}_{c}(\bold{x})=e\sum\limits_{j=1}^{N_{c}}\delta(\bold{x}-\bold{r}^{c}_{j})
\end{equation}
is the microscopic charge density of the counterions,
\begin{equation}
\label{eq:Hp}
H^{(p)}_{e.v.}=\frac{k_{B}Tw}{2}\sum\limits_{i,j=1}^{N_{p}}\int\limits_{0}^{N}\!\!\!\int\limits_{0}^{N}ds_{1}ds_{2}\delta\left(\bold{R}_{i}(s_{1})-\bold{R}_{j}(s_{2})\right)
\end{equation}
is the Hamiltonian of the monomer-monomer excluded volume interaction;
$w=2\pi d^3/3$ is the second virial coefficient of the excluded volume interaction of hard spheres of diameter $d$; $\beta=1/k_{B}T$ is a reciprocal temperature, $k_{B}$ is Boltzmann constant,
\begin{equation}
\label{eq:Hev}
H^{(c)}_{e.v.}=\frac{1}{2}\sum\limits_{i\neq j}V_{hc}(\bold{r}^{c}_{i}-\bold{r}^{c}_{j})
\end{equation} 
is the Hamiltonian of counterion-counterion excluded volume interaction;
\begin{equation}
V_{hc}(\bold{r})=\Biggl\{
\begin{aligned}
\infty,\quad&|\bold{r}|\leq d\,\\
0,\quad& |\bold{r}|> d.
\end{aligned}
\end{equation}
is the hard-core potential,
\begin{equation}
\label{eq:Hev}
H^{(pc)}_{e.v.}=\sum\limits_{i, j}\int\limits_{0}^{N}dsV_{hc}(\bold{r}^{c}_{i}-\bold{R}_{j}(s))
\end{equation}
is the Hamiltonian of counterion-monomer excluded volume interaction.
Moreover, integration measure over the configurations of polymer chains has been introduced:
\begin{equation}
\label{int_conf_pol}
\int d\Gamma_{p}(\cdot)=\int \mathcal{D}\bold{R}_{1}..\int\mathcal{D}\bold{R}_{N_{p}}e^{-\frac{3}{2b^2}\sum\limits_{j=1}^{N_{p}}\int\limits_{0}^{N}ds\dot{\bold{R}}_{j}^{2}(s)}(\cdot)
\end{equation}
where $b$ is the Kuhn length of the segment, $N_{p}$ is the number of polymer chains with $N$
 as the  degree of polymerization;
\begin{equation}
\int d\Gamma_{c}(\cdot)=\Lambda_{c}^{-3N_{c}}\int\limits_{V} d\bold{r}^{c}_{1}..\int\limits_{V} d\bold{r}^{c}_{N_{c}}(\cdot)
\end{equation}
is the integration measure over the phase space of counterions, $N_{c}$ is the total number of counterions, 
in the system of volume $V$. The symbol $\int \mathcal{D}\bold{R}(..)$ in (\ref{int_conf_pol}) 
denotes a functional integration over the polymer chain configurations. Moreover,  
the following normalization condition must hold
\begin{equation}
\int \mathcal{D}\bold{R} e^{-\frac{3}{2b^2}\int\limits_{0}^{N}ds\dot{\bold{R}}^2(s)}=V.
\end{equation}
In order to take into account a monomer-counterion excluded volume interaction we use an assumption 
\begin{equation}
\label{deplit}
\int d\Gamma_{c}\int d\Gamma_{p} e^{-\beta H_{e.v.}^{(pc)}}(\cdot )\rightarrow \int d\Gamma_{c}^{\prime}\int d\Gamma_{p}^{\prime}(\cdot ),
\end{equation}
where 
\begin{equation}
\int d\Gamma_{c}^{\prime}(\cdot)=\Lambda_{c}^{-3N_{c}}\int\limits_{V_{f}} d\bold{r}^{c}_{1}..\int\limits_{V_{f}} d\bold{r}^{c}_{N_{c}}(\cdot),
\end{equation}
and
\begin{equation}
\label{eq:dGamma_p}
\int d\Gamma_{p}^{\prime}(\cdot)=\int^{\prime} \mathcal{D}\bold{R}_{1}..\int^{\prime} \mathcal{D}\bold{R}_{N_{p}} 
e^{-\frac{3}{2b^2}\int\limits_{0}^{N}ds\sum\limits_{i=1}^{N_{p}}\dot{\bold{R}}_{i}^2(s)}(\cdot),
\end{equation}
$v_{mc}=v_{cm}=v_{c}=v_{m}=\pi d^3/6$. The approximation (\ref{deplit}) is a simplest way to include 
into the theory depletion effect \cite{Barrat_Hansen}. The symbol $\int^{\prime}$ in (\ref{eq:dGamma_p}) 
denotes that an integration performed over the free volume for counterions $V_{f}=V-N_{c}v_{mc}$. In the 
renormalized measure $\int d\Gamma_{p}^{\prime}(\cdot)$ we have the following normalization condition
\begin{equation}
\int^{\prime} \mathcal{D}\bold{R} e^{-\frac{3}{2b^2}\int\limits_{0}^{N}ds\dot{\bold{R}}^2(s)}=V_{f}.
\end{equation}

\subsection{Construction of Perturbation Theory of System}
Adopting the notation POCP for the polymer one component plasma and HCOCP for the hard-core one component plasma, 
the model Hamiltonian (\ref{eq:H}) can be rewritten as
\begin{equation}
H=H_{POCP}+H_{HCOCP}+H_{pert},
\end{equation}
where
\begin{equation}
H_{POCP}=H^{(p)}_{e.v.}+\frac{1}{2}\left(\hat{\rho}_{p}V_{c}\hat{\rho}_{p}\right)+(\hat{\rho}_{p}V_{c}\rho_{c})+\frac{1}{2}\left(\rho_{c}V_{c}\rho_{c}\right)
\end{equation}
is a Hamiltonian of polymer one component plasma (POCP),
\begin{equation}
H_{HCOCP}=H^{(c)}_{e.v.}+\frac{1}{2}
\left(\hat{\rho}_{c}V_{c}\hat{\rho}_{c}\right)+(\hat{\rho}_{c}V_{c}\rho_{p})+
\frac{1}{2}\left(\rho_{p}V_{c}\rho_{p}\right)
\end{equation}
is a Hamiltonian of hard-core one component plasma (HCOCP);
\begin{equation}
H_{pert}=\left(\delta\hat{\rho}_{c}V_{c}\delta\hat{\rho}_{p}\right)
\end{equation}
is a perturbation part of the total Hamiltonian; 
$\rho_{c}=eN_{c}/V_{f}$ is an average charge density of counterions, $V_{f}=V-N_{c}v_{cm}$ 
is a free volume for counterions (monomers), $\rho_{p}=-\rho_{c}$ is an average charge density of monomers,
$\delta\hat{\rho}_{p}(\bold{x})=\hat{\rho}_{p}(\bold{x})-\rho_{p}$ is a local charge density fluctuation of monomers,
$\delta\hat{\rho}_{c}(\bold{x})=\hat{\rho}_{c}(\bold{x})-\rho_{c}$ is a local charge density fluctuation of counterions.
We would like to stress that a set of charged flexible polymer chains immersed in a structureless neutralizing background 
we shall call throughout the paper as polymer one component plasma (POCP). Using the formalism of TPT we shall 
use a set of two independent ideal subsystems as a reference system -- POCP of charged macromolecules and HCOCP of counterions.
Using the TPT formalism for the two independent ideal subsystems as a reference system, we obtain
\begin{equation}
Z=Z_{R}\left<e^{-\beta H_{pert}}\right>_{R},
\end{equation}
where symbol $\left<(\cdot )\right>_{R}$ denotes an averaging over microstates of the reference system \cite{Kubo}
\begin{equation}
\left<(\cdot )\right>_{R}=\frac{1}{Z_{R}}\int d\Gamma_{R}e^{-\beta H_{R}}(\cdot ).
\end{equation}
The partition function of the reference system is then given by
\begin{equation}
Z_{R}=\exp\left[-\beta F_{POCP}-\beta F_{HCOCP}\right].
\end{equation}
$H_{R}=H_{HCOCP}+H_{POCP}$ is a Hamiltonian of the reference system.
Thus an expression for the total Helmholtz free energy $F$ within the TPT approach has a following form
\begin{equation}
\label{eq:cum1}
\beta F=\beta F_{POCP}+\beta F_{HCOCP}-\ln\left<e^{-\beta\left(\delta\hat{\rho}_{c}V_{c}\delta\hat{\rho}_{p}\right)}\right>_{R}.
\end{equation}

\subsection{Free Energy of Reference System}

The expression for density of excess free energy of HCOCP has a following form \cite{Brilliantov_HCOCP}
\begin{equation}
\frac{\mathcal{F}_{HCOCP,ex}}{n_{m}k_{B}T}=\frac{4\eta-3\eta^2}{(1-\eta)^2}+
\frac{3}{4}\left[\ln\left(\Theta+c Z_{0}\Gamma_{c}\right)-c\Gamma_{c}\left(3-\frac{2Z_{0}}{\Theta}\right)\right]-\nonumber
\end{equation}
\begin{equation}
\label{eq:F_HCOCP}
-\frac{3}{2}\left(\frac{c\Gamma_{c}Z_{0}}{\Theta}\right)^{3/2}\arctan\left(\sqrt{\frac{\Theta}{c\Gamma_{c}Z_{0}}}\right),
\end{equation}
where $\Gamma_{c}=l_{B}\left(4\pi n_{m}^{\prime}/3\right)^{1/3}$ 
is a plasma parameter, $\eta=\pi d^3 n_{m}^{\prime}/6$ is packing fraction of counterions, 
$n_{m}^{\prime}=n_{m}/\left(1-\frac{\pi d^3n_{m}}{6}\right)$ is a renormalized concentration 
of monomers (counterions), $l_{B}=e^2/\varepsilon  k_{B}T$ is a Bjerrum length,
$n_{m}=N_{p}N/V$ is concentration of monomers (counterions), 
and $c=2/3\left(4/\pi^2\right)^{1/3}$ is a number constant.
Moreover, the following notations have been introduced
\begin{equation}
Z_{0}=\frac{(1-\eta)^4}{1+4\eta+4\eta^2-4\eta^3+\eta^4},
\end{equation}
\begin{equation}
\Theta=1+\frac{6l_{B}}{5d}\frac{\eta^2(1-\eta)^4(16-11\eta+4\eta^2)}{(1+4\eta+4\eta^2-4\eta^3+\eta^4)^2}.
\end{equation}
The equation (\ref{eq:F_HCOCP}) very accurate describes the thermodynamic properties 
of HCOCP in wide ranges of number density and temperature --
from range of applicability of Debye-Hueckel theory to strong coupling regime \cite{Brilliantov_HCOCP}.
Applying an analogous approach to calculation of free energy of POCP 
as well as for HCOCP \cite{Brilliantov_HCOCP},
we arrive at the following expression (see Appendix A):
\begin{equation}
\label{eq:POCP1}
\mathcal{F}_{POCP,ex}=\mathcal{F}_{POCP,e.v.}+\mathcal{F}_{POCP,cor},
\end{equation}
where the correlation contribution into the free energy of POCP can be determined by the expression
\begin{equation}
\label{eq:POCPcor}
\mathcal{F}_{POCP,cor}=-n_{m}k_{B}Tg(\omega ,\sigma),
\end{equation} 
where
\begin{equation}
g(\omega  ,\sigma)=\frac{3}{4}\left(\omega -\ln\left(1+\omega +\frac{\omega }{\sigma^2}\right)\right)+\nonumber
\end{equation}
\begin{equation}
+\frac{3\sqrt{2}}{4}\frac{\omega \left(\sqrt{\omega }\sigma-
\sqrt{\omega  \sigma^2-4}\right)}{\sigma^{3/2}\sqrt{\omega \sigma +
\sqrt{\omega \left(\omega \sigma^2-4\right)}}}
\arctan{\sqrt{\frac{\omega  \sigma^2+\sqrt{\omega \sigma^2\left(\omega \sigma^2-4\right)}}{2\omega}}}+\nonumber
\end{equation}
\begin{equation}
+\frac{3\sqrt{2}}{4}\frac{\omega \left(\sqrt{\omega }\sigma+\sqrt{\omega  \sigma^2-4}\right)}{\sigma^{3/2}\sqrt{\omega \sigma -\sqrt{\omega \left(\omega \sigma^2-4\right)}}}\arctan{\sqrt{\frac{\omega  \sigma^2-\sqrt{\omega \sigma^2\left(\omega \sigma^2-4\right)}}{2\omega }}},
\end{equation}
$\omega =c\Gamma_{m}$, $\Gamma_{m}=\Gamma_{c}=l_{B}\left(\frac{4\pi n_{m}^{\prime}}{3}\right)^{1/3}$ 
is a plasma parameter of monomers, $\sigma=(9\pi^2n_{m}^{\prime})^{1/3}\sqrt{4\pi b^2\alpha/9}$ 
is a dimensionless parameter. The non-electrostatic part $\mathcal{F}_{POCP,e.v.}$ of the excess 
free energy of POCP can be determined by the expression which obtained by Edwards and Muthukumar within a
field-theoretical approach \cite{Edwards_Muthu}:
\begin{equation}
\label{eq:Fv}
\frac{\mathcal{F}_{POCP,e.v.}}{k_{B}T}=\frac{1}{24\pi \xi^3}-\frac{9wn_{m}^{\prime}}{8\pi \alpha  \xi b^2}+\frac{w {n_{m}^{\prime}}^2}{4},
\end{equation}
where correlation length $\xi$ takes the form
\begin{equation}
\label{eq:xi}
\xi=\frac{1}{2}\left(\frac{9}{4\pi n_{m}^{\prime}b^2\alpha}+\sqrt{\left(\frac{9}{4\pi n_{m}^{\prime}b^2\alpha}\right)^2+\frac{2\alpha b^2}{3wn_{m}^{\prime}}}\right),
\end{equation}
and expansion factor $\alpha$ of neutral polymer chain satisfies the following equation
\begin{equation}
\label{eq:alpha}
\alpha^3-\alpha^2=\frac{48w\xi}{\pi b^4}.
\end{equation}

The system of coupled equations (\ref{eq:Fv}), (\ref{eq:xi}) and (\ref{eq:alpha}) determines 
the density of free energy of solution of flexible polymer chains with excluded volume 
and qualitatively describes thermodynamic properties of the solution in the case of good 
solvent from regime of semi-dilute polymer solution 
to concentrated regime \cite{Edwards_Muthu}.

\subsection{Fluctuation Correction to Free Energy of Reference System}
Further, let us calculate the fluctuation correction to the free energy of the reference system at the level of the Gaussian approximation.
Using the standard Hubbard-Stratonovich transformation we rewrite the partition function of the solution in the following form
\begin{equation}
\label{eq:func_int}
Z=Z_{R}\int\frac{\mathcal{D}\Psi}{C}e^{-\frac{1}{2\beta}\left(\Psi,\hat{V}^{-1}\Psi\right)}\left<e^{i\left(\delta\hat{\rho},\Psi\right)}\right>_{R},
\end{equation}
where 
\begin{equation}
\hat{V}\left(\bold{x}-\bold{y}\right)= 
\begin{pmatrix}
0 & V_{c}\left(\bold{x}-\bold{y}\right)\\
V_{c}\left(\bold{x}-\bold{y}\right) & 0\\        
\end{pmatrix}
,
\end{equation}
\begin{equation}
\Psi(\bold{x})=
\begin{pmatrix}
\Psi_{p}(\bold{x})\\
\Psi_{c}(\bold{x})\\     
\end{pmatrix}
\end{equation}
is a vector of auxiliary fields,
\begin{equation}
\delta\hat{\rho}(\bold{x})=
\begin{pmatrix}
\delta\hat{\rho}_{p}(\bold{x})\\
\delta\hat{\rho}_{c}(\bold{x})\\        
\end{pmatrix}
\end{equation}
is a vector of local charge density fluctuations. 
Moreover, the additional notations have been introduced
\begin{equation}
\left(\Psi,\hat{V}^{-1}\Psi\right)=\int d\bold{x}\int d\bold{y}\sum\limits_{\alpha,\delta}\Psi_{\alpha}(\bold{x})\left(\hat{V}^{-1}\right)_{\alpha\delta}\left(\bold{x}-\bold{y}\right)\Psi_{\delta}(\bold{x})
\end{equation}
and
\begin{equation}
\left(\Psi,\delta\hat{\rho}\right)=\int d\bold{x}\sum\limits_{\alpha}\Psi_{\alpha}(\bold{x})\delta\hat{\rho}_{\alpha}(\bold{x}).
\end{equation}
An operator $\hat{V}^{-1}$ can be determined by the following integral relation
\begin{equation}
\int d\bold{z}\sum\limits_{\delta}\left(\hat{V}^{-1}\right)_{\alpha\delta}\left(\bold{x}-\bold{z}\right)\hat{V}_{\delta\gamma}\left(\bold{z}-\bold{y}\right)=\delta_{\alpha\gamma}\delta\left(\bold{x}-\bold{y}\right).
\end{equation}
Performing some cumbersome calculations (see Appendix B) we arrive at the expression
\begin{equation}
\label{eq:F_ex}
\beta \mathcal{F}_{ex}=\beta \mathcal{F}_{POCP,ex}+\beta \mathcal{F}_{HCOCP,ex}+\beta \mathcal{F}_{pert},
\end{equation}
where the perturbation part of the total density of free energy of the solution at the level of Gaussian approximation takes the form
\begin{equation}
\label{eq:pert}
\beta \mathcal{F}_{pert}=\frac{1}{2}\int\frac{d\bold{k}}{(2\pi)^{3}}\ln\left(1-\beta^2\left<\delta\hat{\rho}_{p}(\bold{k})\delta\hat{\rho}_{p}(-\bold{k})\right>_{POCP}
\left<\delta\hat{\rho}_{c}(\bold{k})\delta\hat{\rho}_{c}(-\bold{k})\right>_{HCOCP}\tilde{V}_{c}^{2}(\bold{k})\right),
\end{equation}
where
$\tilde{V}_{c}(\bold{k})=\frac{4\pi}{\varepsilon \bold{k}^2}$ is a Fourier image of Coulomb potential, 
$\left<\delta\hat{\rho}_{p}(\bold{k})\delta\hat{\rho}_{p}(-\bold{k})\right>_{POCP}$ is a Fourier image 
of correlation function of microscopic charge density fluctuations of POCP, and
$\left<\delta\hat{\rho}_{c}(\bold{k})\delta\hat{\rho}_{c}(-\bold{k})\right>_{HCOCP}$ 
is a Fourier image of correlation function of microscopic charge density fluctuations of HCOCP.
The latter has been obtained within the Gaussian approximation in work \cite{Budkov_2013_1} 
and can be written in the following form
\begin{equation}
\label{eq:cor1}
\left<\delta\hat{\rho}_{c}(\bold{k})\delta\hat{\rho}_{c}(-\bold{k})\right>_{HCOCP}=\frac{Z_{0}e^{2}n_{m}^{\prime}\bold{k}^{2}}{\bold{k}^{2}+4\pi n_{m}^{\prime}l_{B}Z_{0}}.
\end{equation}
The corresponding expression for POCP also obtained within Gaussian approximation (see Appendix A) 
has a following form
\begin{equation}
\label{eq:cor2}
\left<\delta\hat{\rho}_{p}(\bold{k})\delta\hat{\rho}_{p}(-\bold{k})\right>_{POCP}=\frac{n_{m}^{\prime}e^{2}S(\bold{k})\bold{k}^{2}}{\bold{k}^{2}+4\pi n_{m}^{\prime}l_{B}S(\bold{k})},
\end{equation}
where $S(\bold{k})=\frac{n_{m}^{\prime}\xi^{3}}{1+\bold{k}^{2}\xi^{2}}$ 
is a well known expression for the structure factor of semi-dilute polymer solution \cite{DeGennes}.

Substituting (\ref{eq:cor1}) and (\ref{eq:cor2}) into (\ref{eq:pert}) and calculating integral we arrive at
\begin{equation}
\label{eq:pert2}
\beta \mathcal{F}_{pert}=\frac{1}{12\pi \xi^{3}}(G(u,v)-G(u,0)),
\end{equation}
where $u=\sqrt{4\pi l_{B}{n_{m}^{\prime}}^{2}\xi^{5}}$, $v=\sqrt{4\pi l_{B}n_{m}^{\prime}Z_{0}}\xi$,
and function $G(u,v)$ has a form
\begin{equation}
\label{eq:G(u,v)}
G(u,v)=1+v^3+\frac{\sqrt{2}\left(2u^2+(v^2+1)\sqrt{(v^2-1)^2-4u^2}-v^4-1\right)\sqrt{v^2+1+\sqrt{(v^2-1)^2-4u^2}}}{4\sqrt{u^2+v^{2}}}+\nonumber
\end{equation}
\begin{equation}
\frac{\sqrt{2}\left(2u^2-(v^2+1)\sqrt{(v^2-1)^2-4u^2}-v^4-1\right)\sqrt{v^2+1-\sqrt{(v^2-1)^2-4u^2}}}{4\sqrt{u^2+v^{2}}}.
\end{equation}

\subsection{Equation of State}
The expression (\ref{eq:F_ex}) in combination with expressions (\ref{eq:F_HCOCP}-\ref{eq:POCP1}) and (\ref{eq:pert2}-\ref{eq:G(u,v)}) defines the excess free energy of salt-free polyelectrolyte solution in the limit of infinitely long polymer chains ($N\rightarrow\infty$) which taking into account the correlation attraction of charged particles, monomer concentration fluctuations in the regime of semi-dilute solution and  mutual influence of the latter effects. It should be noted that the contributions of electrostatic and non-electrostatic interactions are not independent. The osmotic pressure can be obtained by standard way
\begin{equation}
\label{eq:Pi_os}
\Pi=\Pi_{id}+\Pi_{ex},
\end{equation} 
where
\begin{equation}
\Pi_{id}=n_{m}^{\prime}k_{B}T
\end{equation}
is an ideal part of the osmotic pressure and
\begin{equation}
\Pi_{ex}=n_{m}\left(\frac{\partial{\mathcal{F}_{ex}}}{\partial{n_{m}}}\right)_T-\mathcal{F}_{ex}
\end{equation}
is an excess osmotic pressure of the solution.

\section{Numerical results and discussion}
Turning to the numerical calculation we introduce the dimensionless monomer concentration  
$\tilde{n}_{m}=n_{m}d^3$, temperature $\tilde{T}=k_{B}T\varepsilon d/e^2$, 
and the second virial coefficient $\tilde{w}=wd^{-3}$. In addition we assume that $b=d$. Moreover, we determine the dimensionless osmotic 
pressure $\tilde{\Pi}=\Pi\varepsilon d^4/e^2$.

Fig.1 shows the osmotic pressure as a function of monomer concentration for two temperatures $\tilde{T}$. 
At sufficiently low temperatures the appearance of a Van-der-Waals loop indicates a liquid-liquid 
phase separation in the solution. Using the  equality between chemical potentials 
and the osmotic pressures of coexisting phases we obtain a coexistence curve of the liquid-liquid phase 
separation with an upper critical point (see Fig.2). The obtained coexistence curve 
exhibits a highly asymmetric shape, which is in agreement with Monte-Carlo computer simulation results by \cite{Kumar_PRL}. 
Fig.2 also shows for comparison  the values of the critical points obtained within different theories (including the present theory) 
and Monte-Carlo computer simulations. The values of the critical parameters obtained by extrapolation of simulation results 
for $N\rightarrow \infty$ in the work \cite{Kumar_PRL} are: $\tilde{T}_{c}=0.2$ and $\tilde{n}_{m,c}=0.09$. 
The Muthukumar theory \cite{Muthu_2002} sufficiently underestimates
both critical parameters: $\tilde{T}_{c}=\tilde{n}_{m,c}=1/64\pi\approx 0.005$. 
The theory developed by Jiang and co-authors \cite{JIANG_2001} based on MSA, as has already 
been mentioned in the Introduction predicts the critical temperature in close agreement with simulation 
results but underestimates the critical monomer concentration: $\tilde{T}_{c}=0.245$ and $\tilde{n}_{m,c}=0.01$. 
The present theory yields the critical parameters $\tilde{T}_{c}=0.2$ and $\tilde{n}_{m,c}=0.12$ which are in 
close agreement with  Monte-Carlo simulation results.

It is instructive to regard how our theory can describe the thermodynamic properties of polyelectrolyte solution 
in the region above the critical point. Fig. 3 shows the comparison between the dependencies of the osmotic pressure 
on monomer concentration calculated within present theory and within MD simulation at $\tilde{T}=1.2$ in wide range of
monomer concentration \cite{Kremer}. As seen from Fig. 3, our theory gives very good agreement with the results 
of MD simulation. Also our results are in very good agreement with a numerical calculation within PRISM theory 
\cite{Yethiraj_PRISM}.  It should be noted, that two qualitatively different regimes of the 
osmotic pressure behavior take place. At a region of the small monomer concentration where the entropy 
of the mobile counterions gives a basic contribution into the osmotic pressure there is a linear dependence 
on monomer concentration ($\tilde{\Pi}\sim \tilde{n}_m$), whereas in concentrated regime we observe an 
essentially nonlinear behavior ($\tilde{\Pi}\sim {\tilde{n}_m}^{9/4}$). 
The predicted scaling law has been also obtained within a Monte-Carlo simulation \cite{Yethiraj_MC} 
and numerical calculations based on PRISM theory \cite{Yethiraj_PRISM}.

\section{Summary}
We have presented a first-principle statistical theory of salt-free flexible chain 
polyelectrolyte solutions in the regime of a good solvent. The approach is  based on the thermodynamic perturbation 
theory formalism and predicts a liquid-liquid phase separation arising from strong correlation 
attraction of charged particles. Taking into account the correlation attraction of charged particles 
beyond the linear Debye-Hueckel theory and strongly correlated monomer concentration fluctuations 
in the regime of semi-dilute polymer solution we have developed an equation of state for the
salt-free polyelectrolyte solution that accurately predicts the critical 
parameters of the liquid-liquid phase separation and the osmotic pressure in a wide range of 
monomer concentrations in the region above the critical point. We would like to stress that 
our theory provides accurate values of the critical parameters without the concept of charge renormalization 
(e.g. counterion condensation). As is well known, the counterion 
condensation is an essentially nonlinear effect \cite{Manning,Fixman,Dobrynin_2003} that cannot 
be described within the linear Debye-Hueckel type of theory. However, our theoretical model going beyond 
the simple linear theory, implicitly takes into account the effect of charge renormalization.

A limitation of the presented theoretical model is related to the fact that it can be applied only to polyelectrolyte 
solution of sufficiently long polymer chains (i.e., in a case $N\rightarrow \infty$). 
However, as was shown in theoretical works \cite{Muthu_2002,Muthu_2009,JIANG_2001,Budkov_2011,Budkov_PhysA,Budkov_2013_2,Budkov_JPCB} 
and confirmed by Monte-Carlo and MD simulations in references \cite{Kumar_PRL,Kremer,Dobrynin_2003} the degree of polymerization of the polyelectrolyte 
chains has only a weak effect on the thermodynamic properties of salt-free solution. In other words, 
if the degree of polymerization is higher than $\sim 30$, the thermodynamic quantities of polyelectrolyte solution
in the regime of semi-dilute solution only weakly depend on the length of the polymer chains.

The obtained accurate equation of state can be applied to compute thermodynamic properties of 
polyelectrolyte solutions with a low concentration of low-molecular weight 
salts using the ionic strength as a small parameter of perturbation within the thermodynamic perturbation theory \cite{Budkov_2014}. 
In addition, the obtained equation of state can be used to describe inhomogeneous polyelectrolyte 
solutions within the classical density functional theory. Replacing the equation of state of the 
simple one-component plasma \cite{Forsman} by the equation of state into the free energy functional 
instead of allowing to adequately take into account the contribution of the correlation attraction.

In conclusion we would like to speculate about the reasons why a set of two independent one-component plasmas 
can be used for theoretical description of thermodynamics of salt-free polyelectrolyte solution near the critical point. As was predicted in the reference \cite{Mahdi_2000} and confirmed by Monte-Carlo simulation in work \cite{Kumar_PRL}, due to the strong correlation attraction the charged macromolecules near the critical point form a network of polyelectrolyte aggregates. Thus, in this case the charged network can be regarded as a weakly 
fluctuating background neutralizing the charge of counterions, whereas the mobile counterions 
immersed in the polymer network can be regarded as a neutralizing background for charged macromolecules. 
Thus, in order to evaluate the contribution of the remaining effects that are related to the fluctuations 
of charge densities of the neutralizing backgrounds which undergo only slight fluctuations the 
Gaussian approximation can be applied.

In conclusion we would like to estimate the critical parameters in physical units for the real experimental systems. We take the following parametes for the polyelectrolyte and counterions $b\simeq d =0.25~nm$ which approximately corresponding to sodium polystyrene-sulfonate. As an example of polar solvents, where the discussed liquid-liquid phase transition can take place, we take N,N-dimethylformamide ($\varepsilon\simeq 36.7$) and acetonitrile ($\varepsilon \simeq  37.5$). For N,N-dimethylformamide we obtain the following value of the critical temperature $T_{c}\simeq 364~K$. For acetonitrile we get $T_{c}\simeq 356~K$. The critical concentration in both cases $\tilde{n}_{m,c}\simeq 13~M$. Thus this purely electrostatic liquid-liquid phase transition can be realized at experimentally accessible conditions.

\begin{acknowledgments}
This research has received funding from the European Union's Seventh Framework Programme (FP7/2007-2013) under grant agreement No 247500 with project acronym "Biosol". This work was supported by grant from the President of the RF (No MK-2823.2015.3) and by grant from Russian Foundation for Basic Research (No 15-43-03195).
\end{acknowledgments}

\section{Appendix A: Derivation of formulas (\ref{eq:POCPcor}) and (\ref{eq:cor2})} 
In this Appendix we derive the expressions for the excess free energy (\ref{eq:POCPcor}) 
and the correlation function of the charge density fluctuations (\ref{eq:cor2}) of POCP. 

Starting from the partition function of POCP
\begin{equation}
\label{eq:Zscocp}
Z_{POCP}=e^{\beta E_{self}}\int d\Gamma_{p} e^{-\beta H_{POCP,e.v.}-\beta H_{POCP,el}},
\end{equation}
where
\begin{equation}
H_{POCP,e.v.}=\frac{k_{B}Tw}{2}\sum\limits_{i,j=1}^{N_{p}}\int\limits_{0}^{N}\!\!\int\limits_{0}^{N}ds_{1}ds_{2}\delta\left(\bold{R}_{i}(s_{1})-\bold{R}_{j}(s_{2})\right)
\end{equation}
is the monomer-monomer excluded volume interaction part of the total Hamiltonian,
\begin{equation}
\int d\Gamma_{p}(\cdot )=\int \mathcal{D}\bold{R}_{1}..\int\mathcal{D}\bold{R}_{N_{p}}e^{-\frac{3}{2b^2}\sum\limits_{j=1}^{N_{p}}\int\limits_{0}^{N}ds\dot{\bold{R}}_{j}^{2}(s)}(\cdot )
\end{equation}
is the  probability measure over configurations of the Gaussian polymer chains,
\begin{equation}
H_{POCP,el}=\frac{1}{2}\left(\hat{\rho}_{p}V_{c}\hat{\rho}_{p}\right)+\left(\rho_{b}V_{c}\hat{\rho}_{p}\right)+\frac{1}{2}\left(\rho_{b} V_{c}\rho_{b}\right)
\end{equation}
is the  electrostatic interaction part of the total  Hamiltonian due to monomer-monomer,
monomer-background and background-background interaction, $\rho_{b}=-\left<\hat{\rho}_{p}(\bold{x})\right>_{n.ch.}$ 
is the charge density of the neutralizing background, $\beta E_{self}=\frac{\beta NN_{p}e^{2}}{2k_{B}T}V_{c}(0)$ 
is the electrostatic self-energy of monomers and
\begin{equation}
V_{c}(\bold{x}-\bold{y})=\frac{1}{\varepsilon |\bold{x}-\bold{y}|}
\end{equation}
is the Coulomb potential.

We now turn to the calculate the partition function (\ref{eq:Zscocp}) in the 
framework of TPT formalism which was proposed  in the reference \cite{Hubbard} choosing as a reference system a set of neutral polymer chains with excluded volume interactions. As  announced in the main text of the article we are making use of the interpolation formula for the free energy of such system that takes into account the effect of strongly correlated monomer concentration fluctuations in regime of semi-dilute polymer solution \cite{Edwards_Muthu}.

Hence, we obtain
\begin{equation}
\label{eq:Z_R}
Z_{POCP}=Z_{n.ch.}e^{\beta E_{self}}\left<e^{-\frac{\beta}{2} \left(\delta\hat{\rho}_{p}V_{c}\delta\hat{\rho}_{p}\right)}\right>_{n.ch.},
\end{equation}
where $\delta\hat{\rho}_{p}(\bold{x})=\hat{\rho}_{p}(\bold{x})-\left<\hat{\rho}_{p}(\bold{x})\right>_{n.ch.}$ is 
the local charge density fluctuation of the monomers; $Z_{n.ch.}$ is the partition function of a solution of neutral polymer chains. 
The symbol $\left<(\cdot )\right>_{n.ch.}$ denotes averaging over microstates of solution of neutral polymer chains.

Applying to (\ref{eq:Z_R}) the standard Hubbard-Stratonovich transformation we arrive at
\begin{equation}
\label{eq:HS}
Z_{POCP}=Z_{n.ch.}e^{\beta E_{self}}\int \frac{\mathcal{D}\varphi}{C}
e^{-\frac{1}{2\beta}\left(\varphi V_{c}^{-1}\varphi \right)}\left<e^{i\left(\delta\hat{\rho}_{p}\varphi \right)}\right>_{n.ch.},
\end{equation}
where
\begin{equation}
C=\int \mathcal{D}\varphi e^{-\frac{1}{2\beta}\left(\varphi V_{c}^{-1}\varphi \right)}
\end{equation}
is a normalization constant. For sake of simplicity the following short-hand notations 
\begin{equation}
\left(\varphi V_{c}^{-1}\varphi \right)=\int d\bold{x}
\int d\bold{y}\varphi (\bold{x})V_{c}^{-1}(\bold{x}-\bold{y})\varphi (\bold{y})
\end{equation}
and
\begin{equation}
\left(\delta\hat{\rho}_{p}\varphi \right)=\int d\bold{x}\delta\hat{\rho}_{p}(\bold{x})\varphi(\bold{x}).
\end{equation}
have been introduced.
The kernel of the reciprocal operator $V^{-1}_{c}$ can be determined from the relation
\begin{equation}
\int d\bold{z}V_{c}(\bold{x}-\bold{z})V_{c}^{-1}(\bold{z}-\bold{y})=\delta (\bold{x}-\bold{y}).
\end{equation}
Truncating the cumulant expansion in (\ref{eq:HS}) at the second order
\begin{equation}
\label{eq:cum}
\left<e^{i\left(\delta\hat{\rho}_{p}\varphi \right)}\right>_{n.ch.}=\exp\left[i\left(\left<\delta\hat{\rho}_{p}\right>_{c}\varphi \right)-
\frac{1}{2}\left(\varphi\left<\delta\hat{\rho}_{p}\delta\hat{\rho}_{p}\right>_{c} \varphi \right)-..\right],
\end{equation}
we obtain
\begin{equation}
\label{eq:cum3}
Z_{POCP}\approx Z_{n.ch.}e^{\beta E_{self}} \int \frac{\mathcal{D}\varphi}{C}
e^{-\frac{1}{2\beta}\left(\varphi V_{c}^{-1}\varphi \right)-\frac{1}{2}\left(\varphi\left<\delta\hat{\rho}_{p}\delta\hat{\rho}_{p}\right>_{c} \varphi \right)},
\end{equation}
where the symbol $\left<(\cdot )\right>_{c}$ denotes the cumulant average \cite{Kubo}.
Performing the calculation of the Gaussian integral (\ref{eq:cum3}), we arrive at
\begin{equation}
\label{eq:Zpocp}
Z_{POCP}=Z_{n.ch.}\exp\left[\frac{V}{2}\int\frac{d\bold{k}}{(2\pi)^3}\left(\frac{\kappa_{p}^2}{\bold{k}^2}-\ln\left(1+\frac{\kappa_{p}^2}{\bold{k}^2}(1+S(\bold{k}))\right)\right)\right],
\end{equation}
where $\kappa_{p}=\sqrt{4\pi l_{B}n_{m}}$.
We shall use the following approximation for the structure factor of the semi-dilute polymer solution
\begin{equation}
S(\bold{k})\simeq\frac{n_{m}\xi}{\bold{k}^2}\simeq \frac{9}{4\pi\alpha b^2}\frac{1}{\bold{k}^2},
\end{equation}
where the following approximate expression for correlation radius has been used
\begin{equation}
\label{eq:xi2}
\xi\simeq \frac{9}{4\pi n_{m}\alpha b^2}.
\end{equation}
The evaluation (\ref{eq:xi2}) follows from the equation (\ref{eq:xi}) for the case of 
the semi-dilute polymer solution. The expansion factor $\alpha$ of the neutral polymer chain in the case of semi-dilute polymer    solution as follows from (\ref{eq:alpha}) can be evaluated as $\alpha\simeq\left(48/\pi\right)^{1/3}\left(w\xi/b^4\right)^{1/3}$, 
so that evaluation of the correlation radius takes the form
\begin{equation}
\xi \simeq \left(\frac{243}{1024\pi^2}\right)^{1/4}(wb^2)^{-1/4}n_{m}^{-3/4}
\end{equation}
that is in agreement with the well-known scaling result \cite{DeGennes}.

Further, following the idea proposed in the references \cite{Brilliantov_OCP,Brilliantov_HCOCP} 
we introduce an ultraviolet cut-off in the integral over the wave vectors $\bold{k}$ in (\ref{eq:Zpocp}). 
In order to obtain the cut-off parameter $\Lambda$, we assume that total number of collective variables 
$\delta\hat{\rho}_{p}(\bold{k})$ which contribute to the total free energy of the POCP is equal to the total number of degrees of freedom of POCP, i.e. $3NN_{p}$. Therefore, we arrive at the following relation
\begin{equation}
\label{eq:cut-off}
2 V\frac{4\pi}{(2\pi)^3}\int\limits_{0}^{\Lambda}dk k^2 =3NN_{p},
\end{equation}
from which we obtain the expression for cut-off parameter
\begin{equation}
\label{eq:cut-off2}
\Lambda=(9\pi^2 n_{m})^{1/3}.
\end{equation}
The prefactor $2$ on the left-hand side of (\ref{eq:cut-off}) originates from the summation (integration) of
both the real and the imaginary parts of the collective variable $\delta\hat{\rho}_{p}(\bold{k})$.
Thus, using the ultraviolet cut-off (\ref{eq:cut-off}) and taking the integral in (\ref{eq:Zpocp})
we obtain
\begin{equation}
\label{eq:Fscocp}
\mathcal{F}_{POCP}=\mathcal{F}_{POCP,e.v.}+\mathcal{F}_{POCP,cor},
\end{equation}
where
\begin{equation}
\mathcal{F}_{POCP,cor}=-n_{m}k_{B}Tg\left(\omega,\sigma\right)
\end{equation}
is a correlation contribution and
\begin{equation}
\label{eq:g}
g(\omega  ,\sigma)=\frac{3}{4}\left(\omega -\ln\left(1+\omega +\frac{\omega }{\sigma^2}\right)\right)+\nonumber
\end{equation}
\begin{equation}
+\frac{3\sqrt{2}}{4}\frac{\omega \left(\sqrt{\omega }\sigma-\sqrt{\omega  \sigma^2-4}\right)}{\sigma^{3/2}\sqrt{\omega \sigma +\sqrt{\omega \left(\omega \sigma^2-4\right)}}}\arctan{\sqrt{\frac{\omega  \sigma^2+\sqrt{\omega \sigma^2\left(\omega \sigma^2-4\right)}}{2\omega}}}+\nonumber
\end{equation}
\begin{equation}
+\frac{3\sqrt{2}}{4}\frac{\omega \left(\sqrt{\omega }\sigma+\sqrt{\omega  \sigma^2-4}\right)}{\sigma^{3/2}\sqrt{\omega \sigma -\sqrt{\omega \left(\omega \sigma^2-4\right)}}}\arctan{\sqrt{\frac{\omega  \sigma^2-\sqrt{\omega \sigma^2\left(\omega \sigma^2-4\right)}}{2\omega }}},
\end{equation}
$\omega =c\Gamma_{m}$ with $\Gamma_{m}=l_{B}\left(4\pi n_{m}/3\right)^{1/3}$ 
being the plasma parameter of the monomers, $c=2/3\left(4/\pi^2\right)^{1/3}$ a number constant and $\sigma=(9\pi^2n_{m})^{1/3}\sqrt{4\pi\alpha b^2/9}$ is a dimensionless parameter.

Turning to the calculation of the correlation function (\ref{eq:cor2}) of the local charge density of POCP 
at the level of the Gaussian approximation we employ the method of the generating functional \cite{Zinn-Justin}.
The partition function of POCP immersed in an imaginary auxiliary external field $i\psi(\bold{x})$ takes the form
\begin{equation}
Z[\psi]=Z_{n.ch.}e^{\beta E_{self}}\left<e^{-\frac{\beta}{2} \left(\delta\hat{\rho}_{p}V_{c}\delta\hat{\rho}_{p}\right)+i(\psi\delta\hat{\rho}_{p})}\right>_{n.ch.}.
\end{equation}
After introducing the following generating functional
\begin{equation}
\mathcal{G}[\psi]=\frac{Z[\psi]}{Z[0]},
\end{equation}
with $Z[0]=Z_{POCP}$ the correlation function of local charge density of POCP can be determined from 
the double functional derivative of the generating functional $\mathcal{G}[\psi]$, i.e. 
\begin{equation}
\left<\delta\hat{\rho}_{p}(\bold{x})\delta\hat{\rho}_{p}(\bold{y})\right>_{POCP}=-\frac{\delta^{2} \mathcal{G}[\psi]}{\delta\psi(\bold{x})\delta\psi(\bold{y})}\Bigr|_{\psi=0},
\end{equation}
Using the Hubbard-Stratonovich transform of the partition function $Z[\psi]$ we obtain
\begin{equation}
Z[\psi]=Z_{n.ch.}\int\frac{\mathcal{D}\varphi}{C}e^{-\frac{1}{2\beta}(\varphi V_{c}^{-1}\varphi)}\left<e^{i(\varphi\delta\hat{\rho})+i(\psi\delta\hat{\rho})}\right>_{n.ch.}.
\end{equation}
Thus, we obtain the following functional representation of the generating functional
\begin{equation}
\mathcal{G}[\psi]=\frac{1}{Z_{pert}}\int\frac{\mathcal{D}\varphi}{C}e^{-\frac{1}{2\beta}(\varphi V_{c}^{-1}\varphi)}\left<e^{i(\varphi\delta\hat{\rho}_{p})+i(\psi\delta\hat{\rho}_{p})}\right>_{n.ch.},
\end{equation}
where
\begin{equation}
Z_{pert}=\int\frac{\mathcal{D}\varphi}{C}e^{-\frac{1}{2\beta}(\varphi V_{c}^{-1}\varphi)}\left<e^{i(\varphi\delta\hat{\rho}_{p})}\right>_{n.ch.}
\end{equation}
is the perturbation part of the partition function.
Performing a shift of the integration variable $\varphi\rightarrow  \varphi-\psi$ we obtain
\begin{equation}
\mathcal{G}[\psi]=\frac{1}{Z_{pert}}\int\frac{\mathcal{D}\varphi}{C}e^{-\frac{1}{2\beta}(\varphi V_{c}^{-1}\varphi)+\frac{1}{\beta}(\varphi V_{c}^{-1}\psi)-\frac{1}{2\beta}(\psi V_{c}^{-1}\psi)}
\left<e^{i(\varphi\delta\hat{\rho}_{p})}\right>_{n.ch.}.
\end{equation}
Using the cumulant expansion in the integrand
\begin{equation}
\left<e^{i(\varphi\delta\hat{\rho}_{p})}\right>_{R}=e^{i\int d\bold{x}\varphi(\bold{x})\left<\delta\hat{\rho}_{p}(\bold{x})\right>_{c}-
\frac{1}{2}\int d\bold{x}\int d\bold{y}\left<\delta\hat{\rho}_{p}(\bold{x})\delta\hat{\rho}_{p}(\bold{y})\right>_{c}\varphi(\bold{x})\varphi(\bold{y})+..}
\end{equation}
and truncating at the second order we obtain
\begin{equation}
\mathcal{G}[\psi]\simeq\frac{1}{Z_{pert}}\int\frac{\mathcal{D}\varphi}{C}e^{-\frac{1}{2\beta}(\varphi D^{-1}\varphi)+\frac{1}{\beta}(\varphi V_{c}^{-1}\psi)-\frac{1}{2\beta}(\psi V_{c}^{-1}\psi)},
\end{equation}
The kernel of the integral operator $D^{-1}$ can be determined in the following way:
\begin{equation}
D^{-1}(\bold{x}-\bold{y})=V_{c}^{-1}(\bold{x}-\bold{y})+\beta\left<\delta\hat{\rho}_{p}(\bold{x})\delta\hat{\rho}_{p}(\bold{y})\right>_{c}.
\end{equation}

Analogously, truncating the cumulant expansion at the second order in $Z_{pert}$ and taking the Gaussian integral we obtain
\begin{equation}
\mathcal{G}[\psi]\simeq \exp\left[-\frac{1}{2\beta}\left(\psi V_{c}^{-1}\psi\right)+\frac{1}{2\beta}\left(\psi V_{c}^{-1}DV_{c}^{-1}\psi\right)\right]=
\exp\left[-\frac{1}{2\beta}\left(\psi G\psi \right)\right],
\end{equation}
where
\begin{equation}
G=V_{c}^{-1}-V_{c}^{-1}DV_{c}^{-1}=V_{c}^{-1}\left(I-DV_{c}^{-1}\right)=V_{c}^{-1}\left(I-\left(V_{c}^{-1}+A\right)^{-1}V_{c}^{-1}\right)=\nonumber
\end{equation}
\begin{equation}
=V_{c}^{-1}\left(I-\left(I+V_{c}A\right)^{-1}\right)=V_{c}^{-1}\left(I+V_{c}A\right)^{-1}\left(I+V_{c}A-I\right)=\nonumber
\end{equation}
\begin{equation}
\label{eq:G}
=V_{c}^{-1}\left(I+V_{c}A\right)^{-1}V_{c}A=V_{c}^{-1}V_{c}A\left(I+V_{c}A\right)^{-1}=A\left(I+V_{c}A\right)^{-1},
\end{equation}
and $A$ is an integral operator which can be represented as 
\begin{equation}
A(\bold{x}-\bold{y})=\beta\left<\delta\hat{\rho}_{p}(\bold{x})\delta\hat{\rho}_{p}(\bold{y})\right>_{R}= \beta e^2 n_{m}S(\bold{x}-\bold{y}),
\end{equation}
$I$ is the identity operator,
$S(\bold{x}-\bold{y})$ is the structure factor of a neutral polymer chain in good solvent.
Using the relation (\ref{eq:G}) we obtain 
\begin{equation}
\left<\delta\hat{\rho}_{p}(\bold{x})\delta\hat{\rho}_{p}(\bold{y})\right>_{POCP}=\frac{1}{\beta}G(\bold{x}-\bold{y})=\frac{1}{\beta}\int\frac{d\bold{k}}{(2\pi)^3}\frac{\tilde{A}(\bold{k})}{1+\tilde{A}(\bold{k})\tilde{V}_{c}(\bold{k})}e^{i\bold{k}(\bold{x}-\bold{y})},
\end{equation}
where
\begin{equation}
\tilde{A}(\bold{k})=\beta e^2n_{m}S(\bold{k}).
\end{equation}

Thus we arrive at the Fourier transformed correlation function of the charge density fluctuations of POCP
\begin{equation}
\label{eq:corr_rho1}
\left<\delta\hat{\rho}_{p}(\bold{k})\delta\hat{\rho}_{p}(-\bold{k})\right>_{POCP}=
\frac{e^2n_{m}S(\bold{k})}{1+\beta e^2n_{m}\tilde{V}_{c}(\bold{k})S(\bold{k})}.
\end{equation}
Substituting the  $\tilde{V}_{c}(\bold{k})$ into (\ref{eq:corr_rho1}) we arrive at (\ref{eq:cor2}).

\section{Appendix B: Derivation of formula (\ref{eq:pert})}
In this Appendix we present the derivation of the formula  (\ref{eq:pert}) for the perturbation part of the total free energy. 
The partition function of the solution within TPT has the form
\begin{equation}
Z=Z_{R}\int\frac{\mathcal{D}\Psi}{C}e^{-\frac{1}{2}\left(\Psi,\hat{V}^{-1}\Psi\right)}\left<e^{i\left(\delta\hat{\rho},\Psi\right)}\right>_{R}.
\end{equation}
Further we have introduced the following definition for the perturbative part of the partition function
\begin{equation}
Q_{pert}=\frac{Z}{Z_{R}}=\int\frac{\mathcal{D}\Psi}{C}e^{-\frac{1}{2\beta}\left(\Psi,\hat{V}^{-1}\Psi\right)}e^{i\left(\left<\delta\hat{\rho}\right>_{R},\Psi\right)-
\frac{1}{2}(\Psi\left<\delta\hat{\rho}\delta\hat{\rho}\right>_{R}\Psi)}=\nonumber
\end{equation}
\begin{equation}
\label{eq:gauss}
=\int\frac{\mathcal{D}\Psi}{C}e^{-\frac{1}{2}\left(\Psi,\hat{W}^{-1}\Psi\right)},
\end{equation}
where the identity $\left<\delta\hat{\rho}(\bold{x})\right>_{R}=0$, 
and the definition of the inverse operator $\hat{W}^{-1}$
\begin{equation}
\hat{W}_{\alpha\gamma}^{-1}(\bold{x}-\bold{y})=\hat{V}_{\alpha\gamma}^{-1}(\bold{x}-\bold{y})+\beta \left<\delta\hat{\rho}_{\alpha}(\bold{x})\delta\hat{\rho}_{\gamma}(\bold{y})\right>_{R}
\end{equation}
have been taken into account.
Further, calculating the Gaussian integral (\ref{eq:gauss}) we obtain
\begin{equation}
\label{eq:Q_{pert}}
Q_{pert}=\exp\left[\frac{V}{2}\int\frac{d\bold{k}}{(2\pi)^3}\ln{\frac{\det{(\hat{V}^{-1}(\bold{k}))}}{\det{(\hat{W}^{-1}(\bold{k}))}}}\right],
\end{equation}
with the matrices $\hat{W}^{-1}(\bold{k})$ and $\hat{V}^{-1}(\bold{k})$ being
\begin{equation}
\label{eq:matrix1}
\hat{W}^{-1}\left(\bold{k}\right)= 
\begin{pmatrix}
\beta\left<\delta\hat{\rho}_{p}(\bold{k})\delta\hat{\rho}_{p}(-\bold{k})\right>_{POCP} & \frac{1}{\tilde{V}_{c}\left(\bold{k}\right)}\\
\frac{1}{\tilde{V}_{c}\left(\bold{k}\right)} & \beta\left<\delta\hat{\rho}_{c}(\bold{k})\delta\hat{\rho}_{c}(-\bold{k})\right>_{HCOCP}\\        
\end{pmatrix}
,
\end{equation}
\begin{equation}
\label{eq:matrix2}
\hat{V}^{-1}\left(\bold{k}\right)= 
\begin{pmatrix}
0 & \frac{1}{\tilde{V}_{c}\left(\bold{k}\right)}\\
\frac{1}{\tilde{V}_{c}\left(\bold{k}\right)} & 0\\        
\end{pmatrix}
.
\end{equation}
Further, calculating the determinants of the matrices (\ref{eq:matrix1}) and (\ref{eq:matrix2}) and 
substituting results into (\ref{eq:Q_{pert}}) with
$\beta \mathcal{F}_{pert}=-\frac{1}{V}\ln{Q_{pert}}$, we obtain the formula (\ref{eq:pert}).

\begin{figure}[h]
\center{\includegraphics[width=0.7\linewidth]{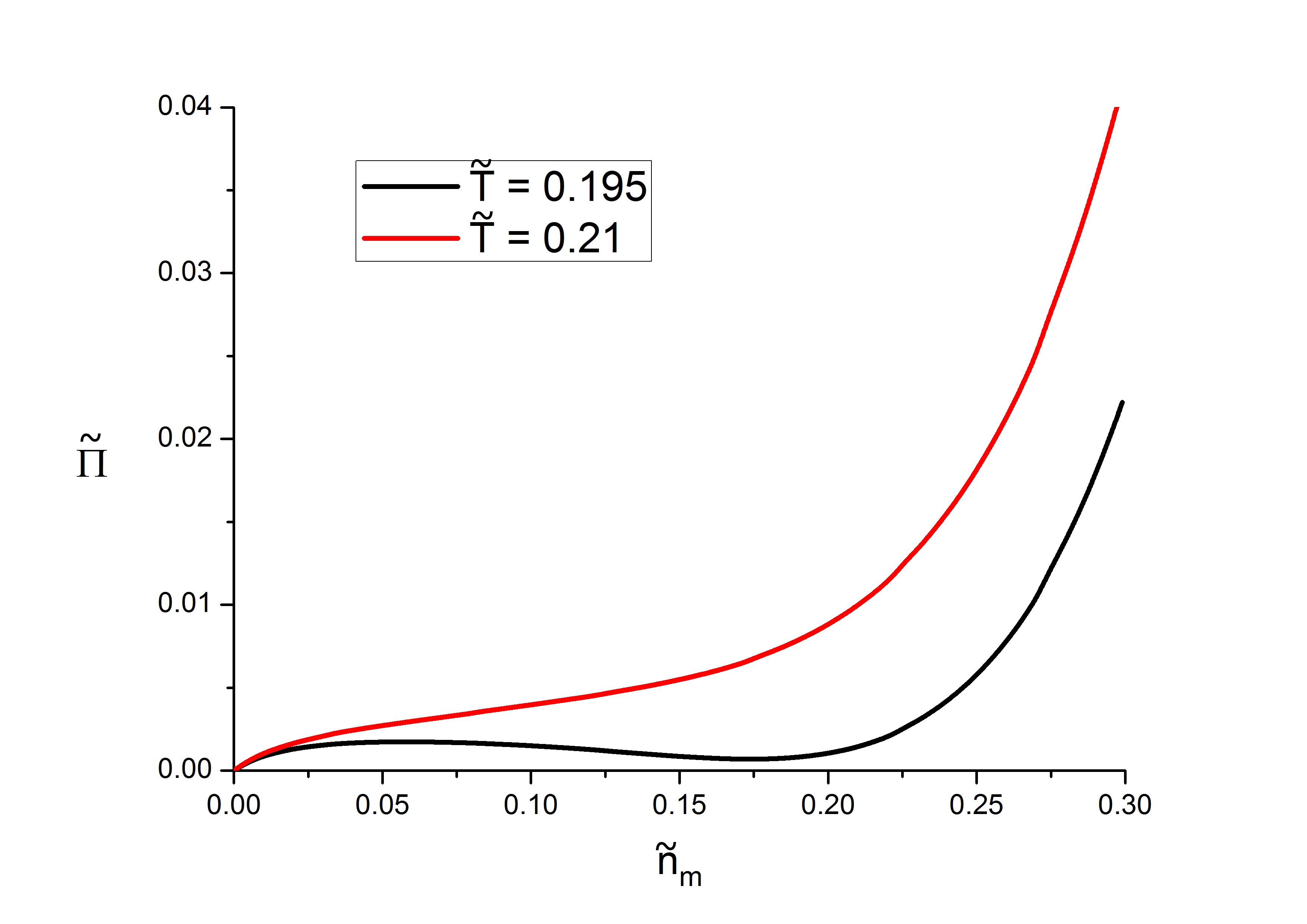}}
\caption{\sl  The osmotic pressure $\tilde{\Pi}$ as a function of the monomer concentration at 
different temperatures $\tilde{T}$. At sufficiently low temperatures the presence of a Van-der-Waals loop 
indicates a liquid-liquid phase separation.}
\label{ris:1}
\end{figure}

\begin{figure}[h]
\center{\includegraphics[width=0.7\linewidth]{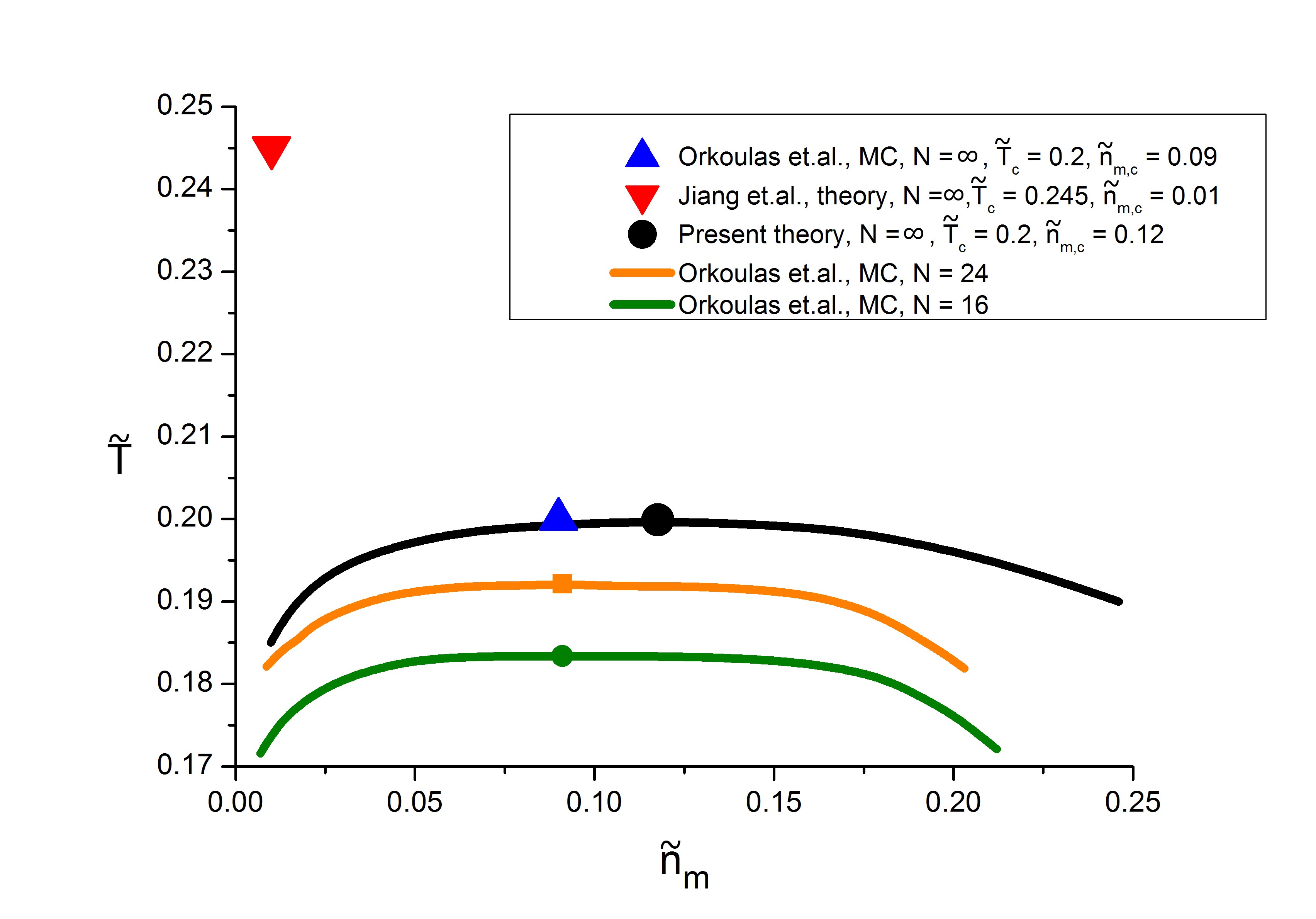}}
\caption{\sl The coexistence curves of salt-free polyelectrolyte solution and critical points that are obtained within present theory and Monte-Carlo simulations \cite{Kumar_PRL} and critical point calculated within theory of Jiang et.al. \cite{JIANG_2001}. The critical point calculated within the present theory is in very good agreement with Monte-Carlo simulations results.}
\label{ris:2}
\end{figure}

\begin{figure}[h]
\center{\includegraphics[width=0.7\linewidth]{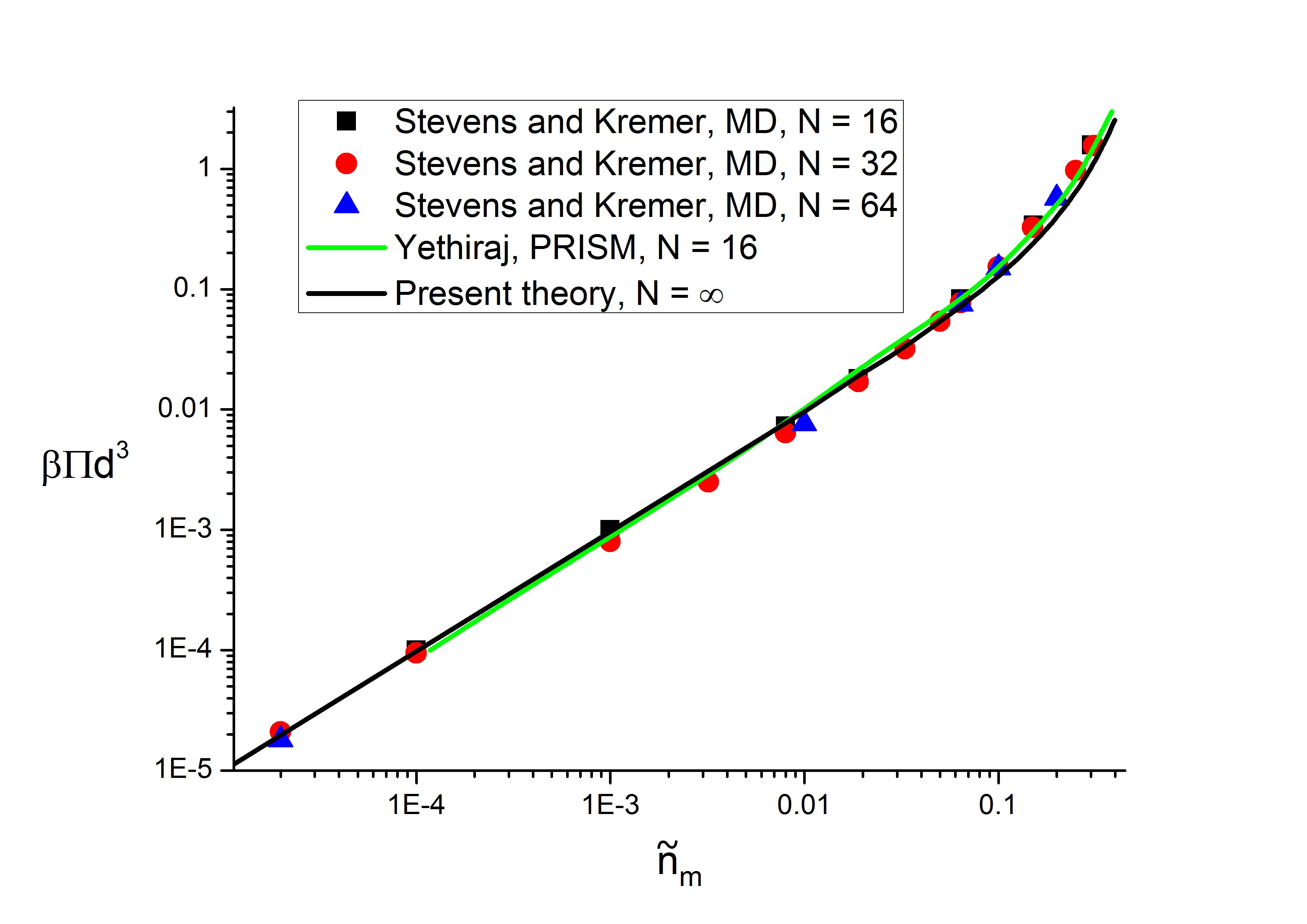}}
\caption{\sl Osmotic pressure $\tilde{\Pi}$ versus monomer concentration $\tilde{n}_{m}$ as obtained 
by the present theory, MD computer simulations \cite{Kremer}, and the PRISM theory \cite{Yethiraj_PRISM} at $\tilde{T}=1.2$.}
\label{ris:3}
\end{figure}


\begin{thebibliography}{99}
\bibitem{Frederickson}
{\sl Fredrickson G.H.} Clarendon Press, Oxford: 2005.

\bibitem{Dobrynin_1}
{\sl A.V. Dobrynin, M. Rubinstein} Prog. Polym. Sci. 2005. V. 30. P. 1049.

\bibitem{Dobrynin_2}
{\sl A.V. Dobrynin} Curr. Opin. Colloid Int. Science. 2008. V. 13. P. 376.

\bibitem{Winkler}
{\sl Winkler R.G., Cherstvy A.G.} Adv Polym Sci (2014) 255, p. 1.

\bibitem{Holm_Review}
{\sl Holm C., Joanny J.F., Kremer K., Netz R.R., Reineker P., Seidel C., Vilgis T.A., Winkler R.G.}
Adv. Polym. Sci. 2004. V. 66.  P. 67.

\bibitem{DeLaCruz_2013}
{Charles E. Sing, Jos W. Zwanikken, and Monica Olvera de la Cruz} Macromolecules 46, 5053 (2013)

\bibitem{Baeurle_2009}
{\sl S.A. Baeurle, M.G. Kiselev, E.S. Makarova, E.A. Nogovitsin} Polymer 50 (2009) 1805-18

\bibitem{Levin}
{\sl Levin Y.} Rep. Prog. Phys. 2002. V. 65. P. 1577.

\bibitem{Khohlov}
{\sl A.Yu. Grosberg and A. R. Khokhlov},  AIP, New York, 1994.

\bibitem{DeGennes}
{\sl de Gennes P.-G.} // Cornell University Press, 1979.

\bibitem{Cherstvy}
{\sl Cherstvy A.G.} J. Phys. Chem. B 2010, 114, 5241.

\bibitem{Borue_1988}
{\sl Borue V.Yu., Erukhimovich I.Ya.} Macromolecules. V. 21. 11. 1988. P. 3240.

\bibitem{Muthu_1996}
{\sl Muthukumar M.} J. Chem. Phys. 105, 5183 (1996).

\bibitem{Joanny_1990}
{\sl Joanny J.F., Leibler L.} J. Phys. (Paris.) 1990. V. 51. P. 545.

\bibitem{Khohlov_1992}
{\sl Khohlov A.R., Nyrkova I.A.} Macromolecules. 1992. V. 25. 5. P. 1493.

\bibitem{Gottschalk_1998}
{\sl Gottschalk M., Linse P., Piculell L.} Macromolecules. 1998. V. 31. P. 8407.

\bibitem{Kramarenko_2002}
{\sl Kramarenko E.Yu., Erukhimovich I.Ya., Khohlov A.R.}  Macromol. Theory Simul. 2002. V. 11. P. 462.

\bibitem{Warren_1997}
{\sl Warren P.B.} ~J. Phys. (Paris). 1997. V. 7. P. 343.

\bibitem{Mahdi_2000}
{\sl Mahdi K.A., Olvera de la Cruz M.}  Macromolecules. 2000. V. 33. P. 7649.

\bibitem{Kumar_PRL}
{\sl Orkoulas G., Kumar S.K.,  Panagiotopoulos A.J.} Phys. Rev. Lett. 2003. V. 90. 4. P. 048303-1.

\bibitem{Ermoshkin}
{\sl  Ermoshkin A.V., Olvera de la Cruz M.} Macromolecules. 2003. V. 36. 20. P. 7824.

\bibitem{Brilliantov_OCP}
{\sl Brilliantov N.V.} Contrib. Plasma Phys. 1998. V.38. 4. P. 489.

\bibitem{Brilliantov_HCOCP}
{\sl Brilliantov N.V., Malinin V.V., Netz R.R.} Eur. Phys. J. D. 2002. V.18. 3. P. 339.

\bibitem{Brilliantov_collaps}
{\sl Brilliantov N.V., Kuznetzov D.V., Klein R.}  Phys. Rev. Lett. 1998. V.81. 7. P. 1433.

\bibitem{Muthu_2002}
{\sl Muthukumar M.} Macromolecules. 2002. V. 35. 24. P. 9142.

\bibitem{Muthu_2009}
{\sl Lee C.-L. and  Muthukumar M.} J. Chem. Phys. 2009. V. 130. P. 024904.

\bibitem{JIANG_2001}
{\sl Jiang J.W., Blum L., Bernard O., Prausnitz J.M.} Mol. Phys. 2001. V. 99. 13. P. 1121.

\bibitem{Barrat_Hansen}
{\sl Barrat J.-L., Hansen J.-P.} University Press, Cambridge: 2003.

\bibitem{Hansen_MacDonald}
{\sl Hansen J. P., Mc Donald I. R.} Academic Press, London; New York; San Francisco: 1976

\bibitem{Edwards}
{\sl Edwards S.F.} Proc. Phys. Soc. 1965. V. 85. P. 613.

\bibitem{Hubbard}
{\sl Hubbard J., Schofield P.}  Phys. Lett. 1972. V. 40A. 3. P. 245.

\bibitem{Edwards_Muthu}
{\sl Muthukumar M., Edwards S.F.} J. Chem. Phys. 1982. V. 76. 5. P. 265.

\bibitem{Budkov_2013_1}
{\sl Budkov Yu. A., Frolov A.I., Kiselev M.G., Brilliantov N.V.} J. Chem. Phys. 2013. V. 139. P. 194901.

\bibitem{Kubo}
{\sl  Kubo R.} J. Phys. Soc. Jap. 1962. V.17. 7. P. 1100.

\bibitem{Manning}
{\sl Manning G.} J. Chem. Phys. 1969. V. 51, P. 924.

\bibitem{Fixman}
{\sl Fixman M.} J. Chem. Phys. 70, 4995 (1979).

\bibitem{Dobrynin_2003}
{\sl Qi Liao, Andrey V. Dobrynin, and Michael Rubinstein} Macromolecules 36, 3399 (2003).

\bibitem{Zinn-Justin}
{\sl Zinn-Justin J.} Clarendon Press, Oxford: 1989.

\bibitem{Kremer}
{\sl Stevens M.J. and Kremer K.} J. Chem. Phys. 103, 1669 (1995).

\bibitem{Yethiraj_MC}
{\sl Chang R. and Yethiraj A.} Macromolecules 2005, 38, 607-616

\bibitem{Yethiraj_PRISM}
{\sl Yethiraj A.} J. Phys. Chem. B, Vol. 113, No. 6, 2009.

\bibitem{Budkov_2014}
{\sl Budkov Yu. A., Kolesnikov A.L., Nogovitsin E.A., Kiselev M.G.} Polymer Science A, 2014, V. 56, 5, p. 697.

\bibitem{Forsman}
{\sl Forsman J., Woodward C.E., Trulsson M.} J. Phys. Chem. B 2011, 115, 4606

\bibitem{Budkov_2011}
{\sl Nogovitsyn E. A., Budkov Yu. A.}  Russ. J. Phys. Chem. A 2011, 85, 1363-1368.

\bibitem{Budkov_PhysA}
{\sl Nogovitsyn E. A., Budkov Yu. A.} Physica A 2012, 391, 2507-2517.

\bibitem{Budkov_2013_2}
{\sl Budkov Yu. A., Nogovitstyn E. A., Kiselev M. G.} Russ. J. Phys. Chem. A 2013, 87, 638644.

\bibitem{Budkov_JPCB}
{\sl Kolesnikov A.L., Budkov Yu.A., Nogovitstyn E.A.} J. Phys. Chem. B 2014, 118, 13037-13049.
\end{thebibliography}
\end{document}